\documentclass[prl,aps,twocolumn,showpacs]{revtex4}

\usepackage{ifpdf}
\ifpdf
\usepackage{subfigure}
\usepackage[pdftex]{graphicx}
\else
\usepackage{graphicx}
\fi

\usepackage{setspace}

\usepackage{psfrag}
\usepackage{epstopdf}
\usepackage{graphicx}
\DeclareGraphicsExtensions{.eps}

\usepackage{amsmath,graphicx,epsfig,bm,color}
\usepackage{amssymb,amsfonts}
\usepackage{rotating}
\usepackage{graphics}
\usepackage{setspace}

\def\R{\mathbb{R}}
\def\Z{\mathbb{Z}}

\begin{document}

\title{\bf Rotation of a Bose-Einstein Condensate held under a toroidal trap}
\author{Amandine Aftalion and Peter Mason,\\
 Ecole Polytechnique, CMAP, UMR-CNRS 7198,\\ F-91128 Palaiseau cedex,
France}

\date{\today}

\begin{abstract}
The aim of this paper is to perform a numerical and analytical study
of a rotating Bose Einstein condensate placed in a harmonic plus
Gaussian trap, following the experiments of \cite{bssd}. The
rotational frequency $\Omega$ has to stay below the trapping
frequency of the harmonic potential and we find that the condensate
has an annular shape containing a triangular vortex lattice. As
$\Omega$ approaches $\omega$, the width of the condensate and the
circulation inside the central hole get large. We are able to provide
analytical estimates of the size of the condensate and the
circulation both in the lowest Landau level limit and the
Thomas-Fermi limit, providing an analysis that is consistent with
experiment.
\end{abstract}

\pacs{03.75.Hh, 05.30.Jp, 67.40.Db, 74.25.Qt} \maketitle

\section*{I. INTRODUCTION}

The investigation of rotating gases or liquids
 is a central issue in the theory of superfluidity
  since they give rise to quantized vortices
\cite{Lifshitz,Donnelly91}.  During recent years, several experiments
 using rotating atomic Bose Einstein condensates have led to the
  observation of vortices. These condensates are usually confined
  in a harmonic potential with
cylindrical symmetry around the rotation axis $z$. Two limiting
regimes occur depending on the ratio of the rotation frequency
$\Omega$ and the trap frequency $\omega$ in the $xy$ plane. When
$\Omega$ is notably smaller than $\omega$, only one or a few
vortices are present at equilibrium \cite{mat,madison}.
 A Thomas Fermi analysis can be performed to analyze this regime
 because the coupling constant  describing the interactions is often large in the
  experiments \cite{Fetter07,Fetter09}.
When $\Omega$ approaches $\omega$, since the centrifugal force
nearly balances the trapping force, the radius of the rotating gas
increases, and the  vortices arrange themselves
 on a triangular lattice
\cite{Ketterle1,Boulder03,Boulder04,Boulder04bis}.
 A new class of phenomena in this regime of fast rotation is predicted in relation with Quantum Hall physics
 \cite{Fetter09,C08,ho,fb,bp,WBP,KCR,abd}.
 Indeed the
one-body Hamiltonian written in the rotating frame is similar to
that of a charged particle in a uniform magnetic field and one can
use the Landau levels structure to analyze the ground state of the
condensate and describe the properties of the lattice.

 In order to analyze the regime of fast rotation, one approach consists
 in
  adding a quartic potential to the harmonic potential. For this type of potential,
   the trapping force is always greater than the centrifugal force so that the regime
    $\Omega>\omega$ can be explored. The condensate is then seen to exhibit a more complex structure
   with regards to its density distribution and the arrangement of vortices
    \cite{ktu,fjs,Aftalion03,abd,FZ,kf,br,cjs}.
    In particular, a multiply quantized vortex, or giant vortex, appears for large
     values of the rotational frequency $\Omega$  and the condensate is
     located within a thin annulus \cite{fjs}. When $\Omega$ is decreased from this
     situation, a circle of vortices exists inside the condensate
     \cite{FZ}.

 A number of experiments have been performed in which a
laser beam is shone into an otherwise harmonically trapped
condensate \cite{bssd,ss,racnhp,wnsbda}, thus creating a trapping
potential of the form harmonic plus Gaussian. Often in experiments,
the laser beam is weak, hence the Gaussian term is small and for the
purpose of analysis can be expanded so that the resulting potential
can be approximated by a harmonic plus quartic potential. A
different approach to analyze these experiments is to consider the
full harmonic plus Gaussian trapping potential.

The aim of this paper is to perform a numerical and analytical study
of a rotating condensate placed in a harmonic plus Gaussian trap as
in the experiment \cite{bssd,ss}.
 The specific feature of the Gaussian potential
 with respect to the quartic one is that the rotation frequency $\Omega$ cannot get
 arbitrarily large but stays below $\omega$, the trapping frequency
 of the harmonic potential.
We will show that according to the parameters of the system, the
 condensate can either be a disk or an annulus. Furthermore we will show that as
 $\Omega$ approaches $\omega$, the condensate always expands to
 become a large annulus with a vortex lattice inside the condensate and a large
 circulation within the central hole. This is in contrast to the harmonic plus quartic
 trap, which develops a giant vortex and a thin annulus. Using the Lowest Landau level (LLL) states,
 we will give an analytical description of the phenomena seen numerically.
 We estimate the radii $R_1$, $R_2$ of the condensate and the
 circulation around the inside hole, of order $R_1R_2$, thus much
 bigger than that given by a uniform lattice (which would be
 $R_1^2$).

This paper is organised as follows. Section II contains a brief
formalisation of the problem, introducing the energy functional
followed by various numerical observations in Sect. III. The
lowest Landau level analysis for the regime $\Omega$ close to
$\omega$ is presented in section IV which provides the main analytical results
of the paper. Finally, section V is devoted to extra computations
 in the
Thomas-Fermi regime.

\section*{II. FORMULATION}

A two-dimensional Bose-Einstein condensate trapped at absolute zero
temperature can be described by a macroscopic condensate wave
function (order parameter) $\Psi$. The ground state of the rotating
system is determined by minimizing the energy functional
$E'=E-\Omega L_z$ where
$L_z=\Psi^*[\hat{z}\cdot(\bm{r}\times\bm{p})]\Psi$ is the $z$
component of angular momentum along the rotation axis (for linear
momentum $\bm{p}$). The energy functional, in the frame rotating
with angular velocity $\Omega$ is then
\begin{equation}
E'=\int_\mathcal{V}\left[\frac{\hbar^2}{2m}|\nabla\Psi|^2+V_{tr}(r)
|\Psi|^2+\frac{U_0}{2}|\Psi|^4-\Omega L_z\right]dV,
\end{equation}
with $r^2=x^2+y^2$ and where the integral is carried out over the
spatial domain $\mathcal{V}$. The trapping potential is composed of
a harmonic plus Gaussian term
\begin{equation}
\label{trap} V_{tr}=V_0\exp{(-2r^2/w_0^2)}+\frac{1}{2}m\omega^2r^2.
\end{equation}
When the atoms are assumed to occupy the ground state of the
harmonic oscillator in the $z$ direction, with energy
$\hbar\omega_z/2$ and extension $a_z=\sqrt{\hbar/m\omega_z}$,
 suppression of the condensate in the $z$ direction is allowed
provided the characteristic energy $\hbar\omega_z$ is very large in
comparison with the other energy scales. Here $\omega_z$ is the
frequency of the confinement in the $z$ direction.
 The
two-dimensional coupling parameter is then
$U_0=\sqrt{8\pi}\hbar^2a_sN/ma_z$  for $N$ identical atoms with
s-wave scatting length $a_s$ \cite{abd}.

The system can be nondimensionalised by choosing $\omega$,
$\hbar\omega$ and $\sqrt{\hbar/(m\omega)}$ as units of frequency,
energy and length respectively. Thus, on defining a non-dimensional
coupling parameter $g=mU_0/\hbar^2$, the energy functional takes the
non-dimensionalised form
\begin{equation}
\label{en}
E'=\int_\mathcal{V}\left[\frac{1}{2}|\nabla\psi|^2+V(r)|\psi|^2+\frac{g}{2}|\psi|^4-\Omega
L_z\right]dV,
\end{equation}
for external toroidal potential trap
\begin{equation}
    \label{trap_toroidal}
    V(r)=Ae^{-l^2r^2}+\frac{1}{2}r^2,
\end{equation}
with $A=V_0/\hbar\omega$ and inverse waist $l=(2\hbar/m\omega
w_0^2)^{1/2}$. The energy functional (\ref{en}) is subject to the
normalization
\begin{equation}
    \label{tor_norm}
    \int_{\mathcal{V}}|\psi|^2r\,dV=1.
\end{equation}

In this scaling, large rotation implies that $\Omega$ gets close to 1.
Note that in experiments $l$ is often small so that the potential $V(r)$  in
Eq. (\ref{trap_toroidal}) can be expanded to give
\begin{equation}
V(r)\sim \frac{1}{2}(1-2Al^2)r^2+\frac 12 Al^4 r^4,
\end{equation}
from which a critical frequency around $1-2Al^2$ is observed
\cite{fjs}. However in this paper we retain the toroidal potential
given by Eq. (\ref{trap_toroidal}) for the numerical and analytical
analysis.

We will perform a full numerical analysis of the experimental case
of \cite{bssd}, which will lead us to a numerical and analytical
description of several model cases which prove to be different from
the harmonic plus quartic trap considered in \cite{fjs,FZ}. In
particular, as $\Omega$ gets close to 1, the condensate has an
annular shape, its width always becomes large and a vortex lattice
is present with a circulation inside the annulus. We are able to
estimate these various quantities.

\section*{III. NUMERICAL OBSERVATIONS}

\subsection{A. The Effective Potential}

When the condensate is put into rotation, the effective trapping
potential to be considered is not given by Eq. (\ref{trap_toroidal}) but is instead given by
\begin{equation}
\label{trap_eff} V_{eff}=V(r)-\frac{1}{2}\Omega^2r^2.
\end{equation}
Therefore according to the values of $A$, $l$ and $\Omega$,
this effective potential can produce either a disk condensate or an
annular condensate.
To see this, notice first that the effective potential (\ref{trap_eff}) has a minimum that occurs
for $r=r_0\ge0$ given by
\begin{equation}
r_0^2=\frac{1}{l^2}\log\left(\frac{2Al^2}{1-\Omega^2}\right),
\end{equation}
provided
\begin{equation}
\label{pot_min} q\equiv\frac{2Al^2}{1-\Omega^2}\ge1.
\end{equation}
  If $q\geq 1$, the effective potential has a local minimum and
it can lead to two different situations, either the condensate is a
disk or an annulus. For existence of an inner boundary, we must have
$q\ge1+\delta$ for some positive (not necessarily small) $\delta$.
As $\Omega\rightarrow1$, $q\ge1+\delta$ is always  satisfied and so
that an inner boundary is  created. The determination of the value
of $\delta$ is not readily obtained as it depends on the
normalisation condition (\ref{tor_norm}). Conversely if $q<1$, then
the condensate is always a disk.

\begin{figure}[ht]
\centering
\includegraphics[scale=0.7]{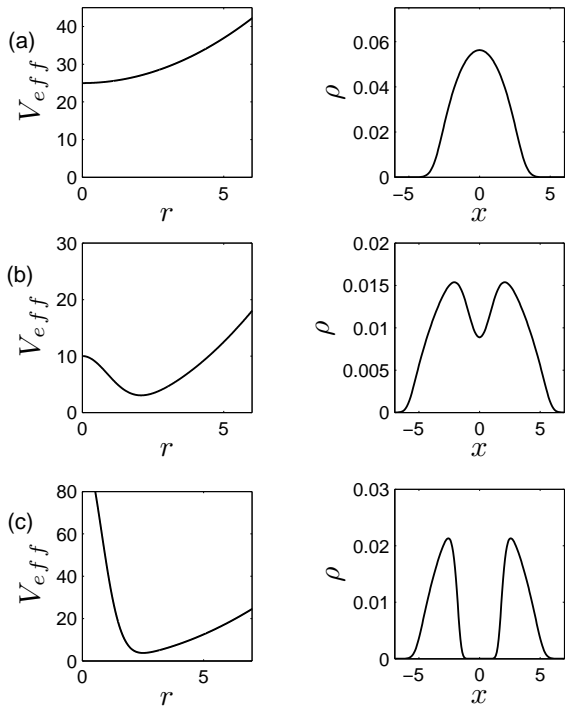}\\
\caption{\baselineskip=-10pt\footnotesize The
effective potential given by Eq. (\ref{trap_eff}) as a function of radial position is shown in the left column. The associated density
($\rho=|\psi|^2$) plots taken along $y=0$ are shown in the right column. Three parameter sets considered are
(a) $\{g, A,l\}=\{100,25,0.03\}$, (b) $\{1000,10,0.75\}$ and (c)
$\{500,100,0.9\}$ all with $\Omega=0$. Distances are measured in
units of $\sqrt{\hbar/(m\omega)}$, density in units of $m\omega/\hbar$ and potentials in units of $\hbar\omega$.}
\label{potential}
\end{figure}

Figure \ref{potential} shows three examples of the effective
potential (\ref{trap_eff}) plotted against radial distance from the
centre of the condensate along constant $\theta$ for the parameters
$\{g, A,l\}=\{100,25,0.03\}$, $\{1000,10,0.75\}$ and
$\{500,100,0.9\}$ with $\Omega=0$. In the first parameter set $q<1$,
the condensate is a disk and the
 density maximum is at the centre.
 In the second parameter set,
 there is a local density minimum at the centre of the condensate, but the condensate
 is still a disk.
The third parameter set displays an inner boundary and the condensate is thus annular.

The effective potential plotted in Fig. \ref{potential} only considers a non-rotating condensate, $\Omega=0$.
 When the condensate is placed under rotation, vortices  form and the shape and size of the condensate
   alter. Numerical simulations on the Gross-Pitaevskii equation are carried out to explore the effect of $\Omega$ on a range of parameter sets for $\{g, A,l \}$. The two-dimensional, dimensionless, Gross-Pitaevskii equation comes directly from the energy functional (\ref{en}) and is
\begin{equation}
i\frac{\partial\psi}{\partial t}=-\frac{1}{2}\nabla^2\psi+(V(r)+g|\psi|^2)\psi-i\Omega\left(y\frac{\partial\psi}{\partial x}-x\frac{\partial\psi}{\partial y}\right),
\end{equation}
for $V(r)$ given by (\ref{trap_toroidal}).
The Gross-Pitaevskii equation is solved numerically in imaginary time (see \cite{cjs,fjs})
by evolving an initial wavefunction for a range of values of $\Omega<1$ in order to find the ground state.
 Three cases of interest, which summarise the numerical results well, are presented below. The three reported parameter sets are; $\{g,A,l\}=\{955.95,24.83,0.07\}$, $\{14,1000,5\}$ and $\{500,60,0.1\}$.

\subsection{B. The Experimental Case of Bretin et al. \cite{bssd}}

\begin{figure}[ht]
\centering
\includegraphics[scale=0.7]{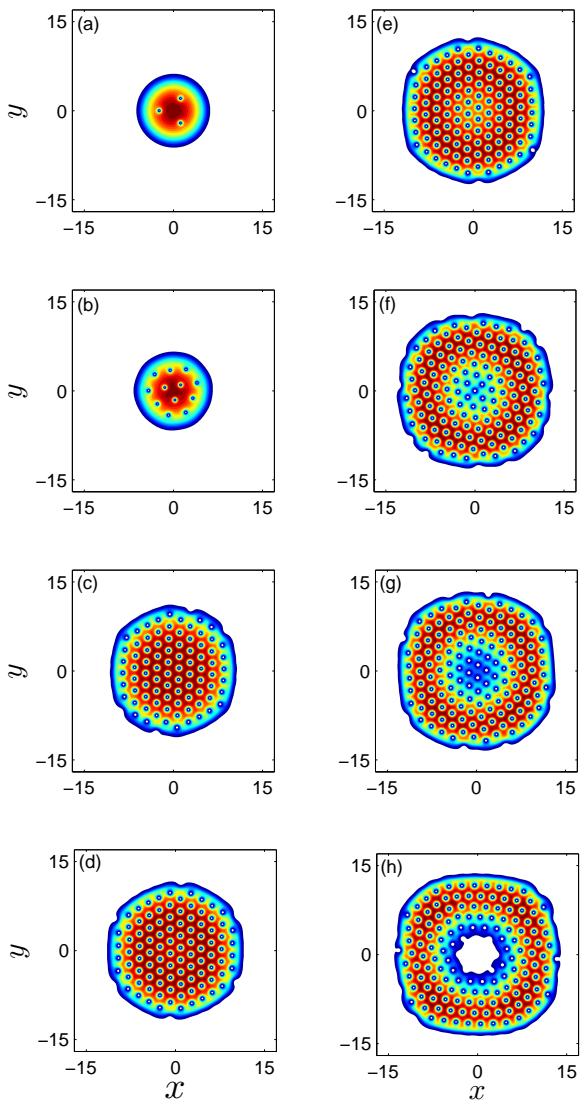}\\
\caption{\baselineskip=-10pt\footnotesize (Color online) Density profiles of a rotating condensate with $\{g,A,l\}=\{955.95,24.83,0.07\}$ corresponding to the (non-dimensionalised) experimental values of \cite{bssd} for (a) $\Omega=0.25$, (b) $\Omega=0.5$, (c) $\Omega=0.874$, (d) $\Omega=0.887$, (e) $\Omega=0.901$, (f) $\Omega=0.914$, (g) $\Omega=0.92$ and (h) $\Omega=0.93$. Distances are measured in units of $\sqrt{\hbar/(m\omega)}$.}
\label{dali2}
\end{figure}

A natural case to numerically simulate is that considered
experimentally by Bretin et al. \cite{bssd} where a harmonically
trapped condensate is put in rotation with a weak laser beam shone
at the origin, modeled by a Gaussian term.  This experimental case
can be described by a two dimensional system as explained in the
introduction, using the dimensional reduction which leads to the
definition of $U_0$. The experimental values of \cite{bssd}
correspond to $\{g,A,l\}=\{955.95,24.83,0.07\}$. A series of contour
plots of the density are shown in Fig. \ref{dali2} (see Fig. 1 of
\cite{bssd}, with the appropriate rescaling of rotational
velocities, so that the $\Omega_{\text{stir}}^{(2)}=60$ of
\cite{bssd} corresponds to $\Omega=0.795$ in this paper and
$\Omega_{\text{stir}}^{(2)}=69$ corresponds to $\Omega=0.914$. The
rotational velocity is calculated from \cite{bssd} using the value
of the frequency in the $x$ and $y$ direction
$\omega_{\perp}^{(0)}/2\pi=75.5Hz$ and not the second stirring phase
frequency $\omega_{\perp}/2\pi=64.8Hz$).

For the slow rotational velocities ($\Omega\lesssim0.5$) of Fig. \ref{dali2}, the condensate is a disk with a small number of vortices (12 vortices are present when $\Omega=0.5$; see Fig. \ref{dali2}(b)) with the vortices forming a triangular
 lattice. As the rotational velocity is increased, the radius of the outer boundary increases while more vortices are accommodated into the condensate. The dynamics here mimic those observed in harmonic traps.

Above some angular velocity the density at the centre of the
condensate  begins to deplete and eventually an inner boundary,
hence an annulus, is created. In the experiments of \cite{bssd}, a
density minimum at the centre first occurred for $\Omega\sim0.874$
(corresponding to $\Omega_{\text{stir}}^{(2)}=66$; see plate (c) of
Fig. 1 in \cite{bssd}).  The numerical simulations here suggest the
onset of the density minimum to be $\Omega\sim 0.887$; see plate (d)
of Fig. \ref{dali2} where the density minimum first appears. Clearer
pictures of the development of the depletion of density at the
centre can be seen in plates (e) and (f).
 It distorts the vortex lattice in much the same way that the
outer boundary does. In Sect IV we note that, under the LLL
approximation, the vortex lattice inside the central hole is
distorted from a regular vortex lattice such that the number of
zeros in the hole is given by $R_1 R_2$, to get a number of order
$R_1^2$, where $R_1, R_2$ are the inner, outer radii.
 Increasing the angular
velocity still further, thus exploring the fast rotation regime
$\Omega\rightarrow1$, details how the density at the centre of the
condensate continues to diminish until for $\Omega\sim0.92$ a
central hole develops (see plate (g)) and the condensate becomes
annular. The central hole grows rapidly; for $\Omega=0.93$, the
central hole is large and there is a circulation equivalent to $11$
vortices (see plate (h)). Our simulations have been carried out up
to $\Omega\sim0.95$. These higher angular velocity simulations
suggest that both the outer and inner radii and also the width of
the condensate increase in size as $\Omega\rightarrow1$.

The experiments of Bretin et al. \cite{bssd} provide  an example of
the transition from a disk condensate with the density maximum at
the center to a disk condensate with a local minimum at the center.
It is reasonably safe to assume that if the angular velocity
 could be further increased in the experiments, the condensate would become
annular, with a large persistent current.  The depletion of density,
which occurs for $\Omega\gtrsim0.887$, creating a distortion in the
vortex lattice and requiring a longer time of convergence for the
numerical simulations, must be one of the reasons that explain the
experimental difficulties in observing the condensate at these
rotation frequencies.

\subsection{C. The Annulus}
\begin{figure}[ht]
\centering
\includegraphics[scale=0.7]{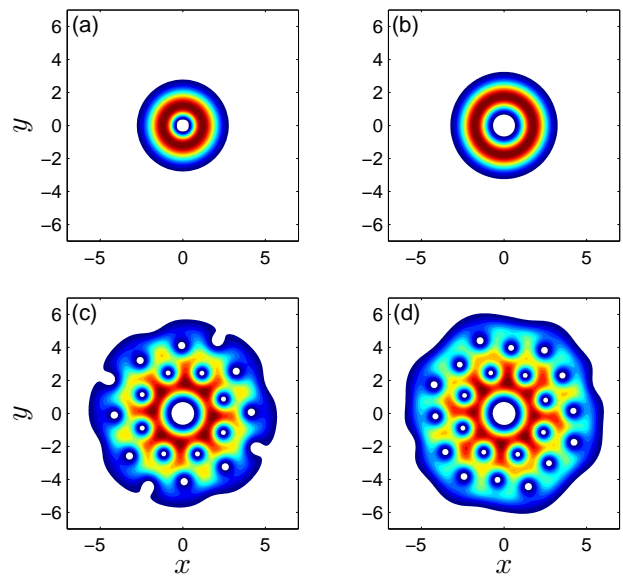}\\
\caption{\baselineskip=-10pt\footnotesize (Color online) Density
profiles of a rotating condensate with
$\{g,A,l\}=\{14,1000,5\}$ for (a) $\Omega=0.25$, (b)
$\Omega=0.9$, (c) $\Omega=0.99$ and (d) $\Omega=0.994$. Distances
are measured in units of $\sqrt{\hbar/(m\omega)}$.} \label{run18}
\end{figure}

Manipulating the values of the parameters $\{g, A, l\}$ can have the effect of altering the shape of the condensate.
 Here we will consider a parameter set that creates
an annular condensate that is present for all $\Omega<1$.
The parameter set is thus chosen to be $\{g,A,l\}=\{14,1000,5\}$.  A selection
of contour plots for various angular velocities are given in Fig.
\ref{run18}. The choice of this parameter set is to illustrate the
behaviour as $\Omega$ gets close to 1.

For low rotational velocities, see Fig. \ref{run18}(a), the
condensate does not contain vortices in the annulus. However the
closer $\Omega$ approaches unity, both the radius of the inner and
outer boundaries increase, but so too does the width of the
condensate. When $\Omega=0.9$ (Fig. \ref{run18}(b)), the condensate
still does not contain any vortices, but for $\Omega=0.99$ (Fig.
\ref{run18}(c)), the condensate  contains two complete rings of
vortices.
For $\Omega=0.99$, there is a multiply quantised vortex at the
centre of the condensate providing a persistent flow with a quantum
of circulation $\nu=3$. The phase profiles also show that there are
further singularities of phase (`invisible' vortices)
 in the outer regions of the condensate.

Note that, for all $\Omega$, the condensate is always an annulus
with the inner and outer radii both increasing as $\Omega$
increases. Furthermore the width of the condensate also
increases so that a thin annulus is never created. The increase in
size of the condensate for $\Omega>0.9$ is marked. This parameter
set explicitly shows the presence, at large $\Omega$, of a central
hole containing circulation together with a vortex lattice in the
bulk of the condensate.

 The choice of this parameter set, especially the value of $g$, is
specifically chosen with reference to the lowest Landau level (LLL)
analysis in Sect. IV. As will noted in Sect. IV, to use the LLL
approximation requires $g(1-\Omega^2)$  to be small. As a
consequence, with an eye on the capability of the numerical
simulations to resolve at  $\Omega$ close to 1, it necessarily
forces $g$ to be not too large, though the main features are
preserved while increasing $g$. We note here that an annular
condensate, existing at all $\Omega$, can be created for a wide
range of values of $g$.

\subsection{D. Density Dip at the Center}

\begin{figure}[ht]
\centering
\includegraphics[scale=0.7]{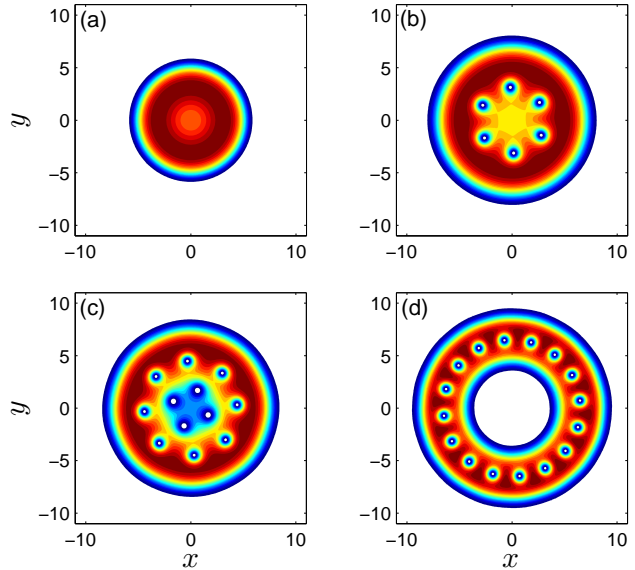}\\
\caption{\baselineskip=-10pt\footnotesize (Color online)
Density profiles of a slow rotating condensate with $\{g,A,l\}=\{500,60,0.1\}$ and (a) $\Omega=0$, (b) $\Omega=0.225$, (c) $\Omega=0.35$ and (d) $\Omega=0.5$. Distances are measured in units of $\sqrt{\hbar/(m\omega)}$.}
\label{run25}
\end{figure}

As a final numerical example,  one can consider the parameter set
$\{g,A,l\}=\{500,60,0.1\}$. The choice of these parameters actually
forces, for $\Omega=0$, the ground state to have a local non-zero
minimum of density at the centre. For these parameters the density
maximum is located at $r=4.2$ when $\Omega=0$ and the condensate is
a disk; a contour plot of the condensate at $\Omega=0$ is given in
Fig. \ref{run25} along with a selection of other plots for various
angular velocities.

Vortices first appear close to the maximum of density instead of
close
 to the center of the condensate. As $\Omega$
 is increased,  the vortex lattice develops close to this initial circle
of vortices. The increase in size of the condensate is visible, as
is the depletion of density at the centre of the condensate, which
turns into a hole: an annulus is formed.

\section*{IV. Lowest Landau Level Analysis}

We turn to the analysis of the ground state of the energy
 (\ref{en}), where $V(r)$ is given by Eq. (\ref{trap_toroidal}) and $\Omega$
tends to 1. We will use a Lowest Landau level analysis
 \cite{ho,abd,ABN2}. Recall that the
spectrum of the Hamiltonian
\begin{equation} H_\Omega =-\frac{1}{2} \nabla^2 +
\frac{1}{2}r^2 -\Omega L_z,
 \label{singlepartH}
 \end{equation} has a Landau
level structure. The lowest Landau level is defined as
\begin{equation}
  \label{eq:LLL3}
  f(x+iy) e^{-\frac \Omega {2}\left(x^2 +
      y^2\right) }, \ f \text{
  analytic}.
\end{equation}
 For such functions, $<H_\Omega \psi,\psi>$ can be
 simplified  (see \cite{abd}) so that
  $E'=\Omega+ E_{LLL}(\psi)$ where
\begin{equation}
E_{LLL}(\psi)=   \int_\mathcal{V}  \left(
  V(r)-\frac 12 \Omega^2 r^2 \right) |\psi|^2+\frac g2 |\psi|^4 dV.
  \label{eq:nrjLLL2}
\end{equation}

The minimization of Eq. (\ref{eq:nrjLLL2})  without the analytic
constraint provides the Thomas-Fermi profile for the coarse grain
density:
\begin{equation}\label{TFF}
|\psi|^2 = \rho_{\rm TF}: = \lambda -\frac 1 g \left ( V(r)-\frac 12
\Omega^2r^2 \right).
\end{equation}
 Such a function can be recovered in the LLL by
the presence of the vortex lattice and this only changes the
coefficient $g$ into $bg$ where $b$ is the Abrikosov parameter as we
explain further below.

We recall that the orthogonal projection for a general
 function $\psi$ onto the LLL
is explicit \cite{B,GiJa} and given by:
\begin{multline}
  \label{eq:projLLL}
  \Pi_{LLL} (\psi) = \frac 1 {\pi} \int
  e^{{-\frac{1}{2}\left(|z|^2 -2 z\overline{z'} +
 |z'|^2\right)}} \psi(x',y')dx'dy',
\end{multline}
where $z=x+i y$ and $z'=x'+i y'$. If an LLL function $\psi$ (i.e
$\psi$ satisfies Eq. \eqref{eq:LLL3}) minimizes the energy
Eq. \eqref{eq:nrjLLL2}, it is a solution of the projected
Gross-Pitaevskii equation:
\begin{equation}
  \label{eq:GPLLL}
  \Pi_{LLL}\left[\left({V(r)-\Omega^2r^2/2}  +
      g|\psi|^2  -\mu\right)\psi \right] = 0,
\end{equation}
where $\mu$ is the chemical potential.

 When $\Omega$ is close to 1, and $V(r)$ is the harmonic plus Gaussian
 potential,
Eq. \eqref{eq:GPLLL} can be approximated by $\Pi_{LLL} \left[ g
|\psi|^2 \psi\right] = \mu \psi,$ which is the equation of the
Abrikosov problem (see \cite{ho,ABN2,Abr}). A  solution  can then be
constructed using the Theta function (see \cite{ABN2} for the
details):
\begin{equation}\label{theta}
\phi(x,y ; \tau) = e^{\frac{1}{2}\left({z^2} - {|z|^2}\right)}
\Theta \left(\sqrt{\frac{\tau_I}{\pi
      }}z, \tau \right),
\end{equation}
where $\tau=\tau_R+i\tau_I$ is the lattice
parameter. The zeroes of the function $\phi $ lie on the lattice
$\sqrt{\frac{\pi}{\tau_I}}\left(\Z\oplus \Z\tau\right)$ and $|\phi|$
is periodic. The optimal lattice, that is the one minimizing
$\mu(\tau)=\int |\phi|^4/(\int |\phi|^2)^2$,
 is hexagonal, which corresponds to $\tau =
e^{2i\pi/3}$ (the integrals are taken on one period).
 For $\tau=e^{2i\pi/3}$, $\mu (\tau)=b\sim 1.16$.

As in \cite{ABN2}, we can construct an approximate ground state by
multiplying the solution, Eq. \eqref{theta}, of the Abrikosov problem by a
profile $\rho$ varying on the same scale as $\rho_{\rm TF}$, which is
large. Since this product is not in the LLL, we project it onto the
LLL and define $v = \Pi_{LLL} \left(\sqrt{ \rho(x,y)}
\phi(x,y;e^{2i\pi/3})\right).$
 Estimating the energy of $v$ yields
 \begin{eqnarray*}
E_{\rm LLL}(v)\sim  \int_{\R^2} \left(
  V(r)-\frac 12\Omega^2 r^2 \right) \rho(x,y)+\frac {g b}2 \rho(x,y)^2.
 \end{eqnarray*}
 This computation assumes
 that
$\phi$ and $\rho$ do not vary on the same
 scale, hence the integrals can be decoupled.  Then, minimizing with respect to $\rho$
 implies that $\rho$ must be a Thomas-Fermi profile (Eq. \eqref{TFF})
 with $g$ changed into $b g$: \begin{equation}\label{TF}
\rho (x,y) = \lambda -\frac 1 {bg} \left ( V(r)-\frac 12 \Omega^2r^2
\right).
\end{equation}
This approximation is valid provided the energy obtained,
 $E_{\rm LLL}(v)$, is much smaller than the gap between two Landau levels
 which is of the order unity.

  In the
case of the harmonic potential $V(r)=r^2/2$, the
 Thomas Fermi profile provides a disk condensate of radius
$R =[4gb/(\pi(1-\Omega^2))]^{1/4}$ which is large when $\Omega$ gets
close to 1. Moreover,  $E_{\rm LLL}(v)$ is of order
$\sqrt{g(1-\Omega^2)}$ which is indeed small when $\Omega$ is close
to 1 and $g$ is not too large, so that the LLL approximation is
satisfied.
In the case of the toroidal potential (\ref{trap_toroidal}), we have
to compute the  Thomas-Fermi profile from Eq. (\ref{TF}) and
discriminate whether it is a disk or an annulus. Then we have to
check whether the LLL approximation is justified, that is whether
$E_{LLL} (v)$ is small.

\subsection*{A. Disc Condensate}
Suppose that the condensate is a disk (recall that this requires that
$q$ is not too large), so that only an outer boundary exists and write
\begin{equation}
\label{tff}
gb|\psi|^2=\mu+\frac{1}{2}(\Omega^2-1)r^2-Ae^{-l^2r^2}.
\end{equation}
To find approximations to the radius of the outer boundary one  can
proceed by taking Eq. (\ref{tff}) and the normalization condition
(\ref{tor_norm}) which provide the necessary starting equations in
order to compute $\mu$. A check on the validity of the LLL approximation is to verify, in the limit $\Omega\rightarrow1$, that the chemical potential $\mu$ is small.

To begin, substitute the density, $|\psi|^2$, from Eq. (\ref{tff}) into
the normalization condition (\ref{tor_norm}) and integrate over the
domain between $0$ and $R_2$, where $R_2$ is defined as the radius
of the outer boundary. It follows that
\begin{equation}
\label{eqqq}
\frac{gb}{\pi}=\mu R_2^2+\frac{1}{4}(\Omega^2-1)R_2^4+\frac{A}{l^2}(e^{-l^2R_2^2}-1),
\end{equation}
which explicitly contains the chemical potential $\mu$. To  remove
$\mu$ from the calculations, one can note from Eq. (\ref{tff}) that
\begin{equation}
\label{eq211}
\frac{\mu-V_{eff}(r)}{gb}\bigg|_{R_2}=|\psi|^2\big|_{R_2}=0,
\end{equation}
from which
\begin{equation}
\label{eq311} \mu=Ae^{-l^2R_2^2}-\frac{1}{2}(\Omega^2-1)R_2^2.
\end{equation}
Using Eq. (\ref{eq311}) in Eq. (\ref{eqqq}) gives
\begin{equation}
\label{wwww}
\frac{gb}{\pi}=\frac{1}{4}(1-\Omega^2)R_2^4+A\left[R_2^2e^{-l^2R_2^2}+\frac{1}{l^2}(e^{-l^2R_2^2}-1)\right].
\end{equation}
At this stage we introduce the parameter $p$, defined as
\begin{equation}
\label{eqp} p\equiv\frac{gl^4}{\pi(1-\Omega^2)}.
\end{equation}
We are interested in the limit $\Omega$ close to
1, i.e.  $p\gg1$, which corresponds to $l^2R_2^2\gg1$. Then the exponential
terms in Eq. (\ref{wwww}) immediately disappear and it follows that
\begin{equation}
\label{rr2}
R_2\sim\left(\frac{4}{(1-\Omega^2)}\left[\frac{gb}{\pi}+\frac{A}{l^2}\right]\right)^{1/4}
\end{equation}
and
\begin{equation}
\label{mm}
\mu\sim\left((1-\Omega^2)\left[\frac{gb}{\pi}+\frac{A}{l^2}\right]\right)^{1/2}.\\
\end{equation}
This implies that  $l^2R_2^2$ is large and $\mu$ is small since
 $\Omega$ is close to 1.
Recall that the assumption of a disk condensate requires $q$ to be
bounded, $q<1+\delta$, and thus (as $\delta$ is not large)
\begin{equation}
\label{al2}
Al^2<\frac{(1+\delta)(1-\Omega^2)}{2},
\end{equation}
so that the ratio $Al^2$ has to be small in this regime.

\subsection*{B. Annular Condensate}

Expressions (\ref{rr2})-(\ref{mm}) are valid strictly when there is
no inner boundary which occurs when $q<1+\delta$. If $q\ge1+\delta$
then an inner boundary, at $r=R_1$,  develops.

One again starts from Eq. (\ref{tff}), but this time the integration is taken
over the domain between $R_1$ and $R_2$. It then follows that
\begin{equation}
\label{eq1}
\frac{gb}{\pi}=\mu(R_2^2-R_1^2)+\frac{1}{4}(\Omega^2-1)(R_2^4-R_1^4)+\frac{A}{l^2}\left(e^{-l^2R_2^2}-e^{-l^2R_1^2}\right),
\end{equation}
which again explicitly contains the chemical potential $\mu$. To remove
$\mu$ from the calculations,  we note from Eq. (\ref{tff}) that
\begin{equation}
\label{eq2}
\frac{\mu-V_{eff}(r)}{gb}\bigg|_{R_1,R_2}=|\psi|^2\big|_{R_1,R_2}=0,
\end{equation}
from which we recover Eq. (\ref{eq311})
and
\begin{equation}
\label{eq4}
e^{-l^2R_2^2}-e^{-l^2R_1^2}=\frac{1}{2A}(\Omega^2-1)(R_2^2-R_1^2).
\end{equation}
Upon substitution of Eq. (\ref{eq311}) and Eq. (\ref{eq4}) into Eq. (\ref{eq1}) and writing the `area' of the annulus as $X\equiv R_2^2-R_1^2$, we get
\begin{equation}
\label{nonu} gb=\frac{1}{2} \pi
    X(\Omega^2-1)\left[\frac{Xe^{-l^2X}}{e^{-l^2X}-1}+\frac{1}{l^2}-\frac{X}{2}\right].
\end{equation}
Since $p$ is large (because $\Omega$ is close to 1), we can assume
that $l^2X$ is large so that the first term in the squared brackets
of Eq. (\ref{nonu}) is negligible, which leaves
\begin{equation}
gb=\frac{1}{2}  \pi
    X^2(\Omega^2-1)\left[\frac{1}{l^2X}-\frac{1}{2}\right].
\end{equation}
Now, as $l^2X$ is taken to be large, one can neglect $1/l^2X$ in front of the factor of a half in the square bracket. Formally, this corresponds to the following being satisfied
\begin{equation}
\frac{1}{1-\Omega^2}\gg\frac{\pi}{gl^4}.
\end{equation}
Thus one obtains an expression for $X$,
\begin{equation}
X\sim\sqrt{\frac{4gb}{\pi(1-\Omega^2)}}, \label{ds}
\end{equation}
so that $l^2X$ is indeed large when $\Omega$ gets close to 1.
Furthermore, from (\ref{eq4})
\begin{equation}
e^{-l^2R_1^2}=\frac{(\Omega^2-1)X}{2A(e^{-l^2X}-1)},
\end{equation}
which, on using the derived approximation for $X$, Eq. (\ref{ds}), implies that
\begin{equation}
\label{r1large}
R_{1}\sim\frac{1}{l}\left[\ln\left(A\sqrt{\frac{\pi}{gb(1-\Omega^2)}}\right)\right]^{1/2},
\end{equation}
giving an expression for the radius of the inner boundary. The value of the chemical potential can then easily be found by taking Eq. (\ref{eq2}) evaluated at $R_1$, with the expression for $R_1$ following directly from Eq. (\ref{r1large});
\begin{eqnarray}
\mu&=&\frac{1}{2}(1-\Omega^2)R_1^2+Ae^{-l^2R_1^2}\nonumber\\
&\sim&\sqrt{\frac{gb(1-\Omega^2)}{\pi}}, \label{chem}
\end{eqnarray}
provided
\begin{equation}
\label{appx1}
\ln\left(A\sqrt{\frac{\pi}{gb(1-\Omega^2)}}\right)\ll\left(\frac{4gbl^4}{\pi(1-\Omega^2)}\right)^{1/4}.
\end{equation}
In the limit $\Omega\rightarrow1$, we see indeed that $\mu\rightarrow0$,
 thus justifying the LLL approximation provided $g$ is not too large. The radius of the outer boundary can be
found by taking Eq. (\ref{eq311}) with $\mu$ given by (\ref{chem}).
Then
\begin{equation}
\label{r2large}
R_2\sim\left(\frac{4gb}{\pi(1-\Omega^2)}\right)^{1/4},
\end{equation}
which is just $\sqrt X$, implicitly implying through $(\ref{appx1})$
that $R_{2}^2\gg R_{1}^2$.

To find the width of the condensate, $d=R_{2}-R_{1}$, it is necessary that $R_{1}$ be neglected in front of $R_{2}$, i.e. that $R_{1}/R_{2}\ll 1$ which again is equivalent to Eq. (\ref{appx1}) being satisfied.
Thus, the width of the condensate $d$ is found from
\begin{eqnarray}
X&=&(R_2-R_1)(R_1+R_2)\nonumber\\
\Rightarrow
d&=&\frac{X}{R_{1}+R_{2}}\sim\frac{X}{R_{2}}=\left(\frac{2gb}{\pi(1-\Omega^2)}\right)^{1/4}.
\end{eqnarray}
Notice that both the width of the condensate $d$ and the inner $R_1$
 and outer boundaries $R_2$ get larger as $\Omega$ increases and eventually tend to infinity as $\Omega\rightarrow1$. However, while the width of the condensate and the outer boundary grow like $(1-\Omega^2)^{-1/4}$, the inner boundary grows like $\ln[(1-\Omega^2)^{-1/2}]^{1/2}$. Thus, in the limit $\Omega\rightarrow1$, the condensate forms an infinitely thick annulus with both the inner and outer boundary tending to infinity and with an infinitely large central hole.

The Cauchy formula allows us to compute the number of zeroes
$\nu(R_1)$ inside the disk of radius $R_1$:
\begin{equation}\label{eq:Cauchy}
\nu(R_1)=\frac{R_1 R_2}{2\pi }\int_0^{2\pi}\!\!\!\! d\theta
\frac{\int e^{\frac
{R_1}{R_2}\overline{z'}}e^{-|z'|^{2}}\overline{z'}\sqrt{\rho(z')}
\phi (z'e^{-i\theta}, \tau)dz'}{\int e^{\frac
{R_1}{R_2}\overline{z'}}e^{-|z'|^{2}}\sqrt{\rho(z')} \phi
(z'e^{-i\theta},\tau)dz'},
\end{equation}
where $\phi$ comes from (\ref{theta}). Using a Laplace method to
evaluate the integrals, we see that for large $R_1$, since $R_1/R_2$
 tends to 0,
\begin{equation}
\label{n} \nu(R_1)\sim R_1 R_2.
\end{equation}
 Note that a regular lattice in a disk of radius $R_1$ would give
$\nu(R_1)\sim R_1^2$, which is much smaller.

\subsection*{C. Summary of the main results}
We have derived that as $\Omega$ approaches 1, the condensate has
 an annular shape with a triangular vortex lattice inside.
  Eq.'s (\ref{r1large}), (\ref{r2large}), (\ref{chem}) and
(\ref{n}) provide the inner, outer radii of the condensate, the
chemical potential and the circulation in the inside hole. In the
LLL approximation, we must have $\sqrt{g(1-\Omega^2)}\ll1$.
Therefore in the limit $\Omega\rightarrow1$, the parameter set
$\{g,A,l\}=\{14,1000,5\}$, described in Set. IIIC and with a series
of contour plots (Fig. \ref{run18}) showing that the condensate is
always annular, satisfies this condition.  Table \ref{run18t} gives
a comparison between the numerical and analytical values for this
parameter set and provide a good  check on the estimates derived in
Eq.'s (\ref{r1large}), (\ref{chem}), (\ref{r2large}) and (\ref{n}).

\begin{table}
\caption{\label{run18t}A comparison, for parameters
$\{g,A,l\}=\{14,1000,5\}$, of the values of the inner radius $R_1$,
outer radius $R_2$ and quantum of circulation $\nu$ for fast
rotation calculated in the LLL and given by Eq.'s (\ref{r1large}),
(\ref{r2large}) and (\ref{n}) respectively. Numerical values are
provided (subscripts $n$) as a comparison. The chemical potential
$\mu$ is also given, calculated from Eq. (\ref{chem}).}
\begin{ruledtabular}
\begin{tabular}{lccccccc}
$\Omega$&$R_1$&$R_{1n}$&$R_{2}$&$R_{2n}$&$\nu$&$\nu_n$&$\mu$\\
\hline
0.9    & 0.53 & 0.66 & 3.23 & 3.25 & 1.70 & 2 & 0.99\\
0.99   & 0.57 & 0.69 & 5.68 & 5.71 & 3.22 & 3 & 0.32\\
0.994  & 0.58 & 0.70 & 6.45 & 6.38 & 3.72 & 3 & 0.25\\
\end{tabular}
\end{ruledtabular}
\end{table}

\section*{V. Thomas-Fermi approximation}

In the case of the experiments of Bretin et al. \cite{bssd},
$\Omega$ is not very close to 1 and $g$ is large, so the LLL
analysis of Sect IV does not adequately describe this experiment. In
order to describe it better, and since $g$ is large, we can use the
Thomas-Fermi (TF) approximation. In the TF approximation, the vortex
cores are a small perturbation with respect to the density profile
and yield similar equations to those previously derived in Sect IV,
except that now there is no factor $b$ entering the TF density
profile of Eq. (\ref{tff}) since the vortex cores are small.  These
computations are similar to those in
 \cite{fjs}, however due to the Gaussian trapping potential used here,
   the condensate becomes a thick annulus, with many vortices, instead of a thin
   annulus with no vortex as in \cite{fjs}. This changes
   significantly the approximations and requires the analysis of
   several cases according to the magnitude of $l^2R^2$, $R$ being
   the
    radius of the condensate.

\subsection*{A. Disk condensate}

If there is no inner boundary then the Thomas-Fermi
approximation leads to
\begin{equation}
\label{tf} g|\psi|^2=\mu+\frac{1}{2}(\Omega^2-1)r^2-Ae^{-l^2r^2}.
\end{equation}
We substitute the density $|\psi|^2$ from Eq. (\ref{tf}) into the
normalization condition (\ref{tor_norm}) and integrate over the
domain between $0$ and $R_2$, where as before $R_2$ is defined as the radius
of the outer boundary. It follows that
\begin{equation}
\label{eqq} \frac{g}{\pi}=\mu
R_2^2+\frac{1}{4}(\Omega^2-1)R_2^4+\frac{A}{l^2}(e^{-l^2R_2^2}-1),
\end{equation}
which explicitly contains the chemical potential $\mu$. To remove
$\mu$ from the calculations, one can note from Eq. (\ref{tf}) that
\begin{equation}
\label{eq21}
\frac{\mu-V_{eff}(r)}{g}\bigg|_{R_2}=|\psi|^2\big|_{R_2}=0,
\end{equation}
from which we get
\begin{equation}
\label{eq31} \mu=Ae^{-l^2R_2^2}-\frac{1}{2}(\Omega^2-1)R_2^2,
\end{equation} which is the same as (\ref{eq311}).
Using Eq. (\ref{eq31}) in Eq. (\ref{eqq}) yields
\begin{equation}
\label{www}
\frac{g}{\pi}=\frac{1}{4}(1-\Omega^2)R_2^4+A\left[R_2^2e^{-l^2R_2^2}+\frac{1}{l^2}(e^{-l^2R_2^2}-1)\right].
\end{equation} If $l$ is taken to be small as in the experiments, we can
assume that $p$ is small and thus that $l^2R_2^2$ is small. Then the
 exponentials in $l^2R_2^2$  can be
expanded and it readily follows that
\begin{equation}
\label{r222}
p\sim\frac{1}{4}l^4R_2^4(1-q)+\frac{1}{6}ql^6R_2^6,
\end{equation}
where we have used the expressions for $q$ and $p$
 (Eq.'s (\ref{pot_min}) and (\ref{eqp}) respectively).
 Equation (\ref{r222}) is a cubic equation in $l^2R_2^2$.
 When $q<1$, the last term in Eq. (\ref{r222}) can be neglected and we get
 \begin{equation}\label{R2appTF}R_2^4\sim \frac{4p}{l^4(1-q)}=
 \frac {4g}{\pi ((1-\Omega^2)-Al^2)}.\end{equation} When $q$ becomes
 larger than 1,
 then $1-q<0$, but since we assume that we have a disk condensate, we
recall that $1-q$ is small. In such situations, the last term in Eq.
(\ref{r222}) cannot be neglected and is crucial to get the correct
sign on the right hand side of (\ref{r222}).

  The chemical potential is derived
 from Eq. (\ref{eq31})
\begin{equation}
\label{muuu} \mu\sim
A\left(1-\frac{p}{q}\right)+\frac{1}{2}R_2^2(1-\Omega^2)(1-q)+\frac{1}{4q}Al^4R_2^4(1+q),
\end{equation}
with the value of $R_2$ obtained from Eq. (\ref{r222}). This allows
us to check that $g|\psi(0)|^2\sim\mu-A$ is large, provided
$1-\Omega^2$ is not too small,  $g$ is large, $l$ is small so that
$p$ is small. Hence, this justifies our use here of the TF
approximation. This is in particular the case for the parameters
$\{g,A,l\}=\{955.95,24.83,0.07\}$ corresponding to the experiments
of \cite{bssd}.

\begin{table}
\caption{\label{dalit}A comparison, for parameters
$\{g,A,l\}=\{955.95,24.83,0.07\}$, for $\Omega$ such that the
condensate is a disk. The value of the radius of the condensate for
different rotational velocities calculated asymptotically from Eq.
(\ref{r222}) under the assumption of the TF approximation and
calculated numerically is shown.}
\begin{ruledtabular}
\begin{tabular}{lcc}
$\Omega$&$R_2$ (asymptotically)&$R_2$ (numerically)\\
\hline
0.1   & 6.29  & 6.18\\
0.25  & 6.40  & 6.19\\
0.5   & 6.87  & 6.70\\
0.795 & 8.97  & 9.00\\
0.821 & 9.44  & 9.49\\
\end{tabular}
\end{ruledtabular}
\end{table}
Table \ref{dalit} gives a comparison between the numerical values
and analytical values (calculated from Eq. (\ref{r222})) for the
boundary of the condensate when the condensate is still a disk
($\Omega\lesssim0.92$), with the agreement found to be extremely
good.

\hfill

A specific feature of our numerics in the case
$\{g,A,l\}=\{500,60,0.1\}$ is to find that for certain values of the
parameters, since the maximum of the density is not at the origin,
that vortices  appear on a specific circle rather than at the origin
(see Figure \ref{run25}(b)).
 We call $\rho (r)$ the  average of $|\psi|^2$ on a circle of radius
  $r$. Then $\rho$ is not far from the TF approximation of $|\psi|^2$
   given by (\ref{tf}), but not exactly since the presence of the circle of vortices
    has an influence on its
shape. A computation similar to that in \cite{FZ} would be required
 to determine $\rho$ in this setting. Once this is done, one should
 be able to use the results of \cite{A}, chapter 3, which states
 that the radius where the circle of vortices appear is given by
 the location where the function
$\zeta /\rho$ reaches its maximum, where $\zeta (r)=\int_r^R
s\rho(s)\ ds$, $R$ being the radius of the condensate.  This follows
from an expansion of the Gross Pitaevskii energy, where the leading
order is given by the energy of vortices of order $\rho(r) \ln \xi$,
where $\xi$ is the scattering length, minus the $L_z$ term, which
can be estimated as $-\Omega \zeta (r)$. Thus, the lowest $\Omega$
for nucleation of vortices is achieved when there is a  radius where
$\zeta /\rho$ is minimal.

\subsection*{B. Annular condensate}

When the condensate is an annulus, then in the TF approximation, we
need to take into account the quantum of circulation $\nu$ in the
inner hole of the condensate. We assume that $\psi=|\psi|e^{i\nu
S}$,  where $S$ is the phase. Note that this is not needed in the
LLL approach,  because the circulation is incorporated into the LLL
wave function and we can compute it directly from (\ref{eq:Cauchy}).

 The TF density expression (\ref{tf}) is
adjusted to
\begin{equation}
\label{tfnu}
g|\psi|^2=\tilde{\mu}+\frac{1}{2}\left[(\Omega^2-1)r^2-\frac{\nu^2}{r^2}\right]-Ae^{-l^2r^2},
\end{equation}
where $\tilde{\mu}=\mu+\Omega\nu$. The TF density (\ref{tfnu}) and
the normalisation condition (\ref{tor_norm}) provide the starting
points from which approximations to the values of the radii of the
condensate boundaries and thus the width of the condensate can be
found. It follows then that
\begin{equation}
\label{gg}
0=\tilde{\mu}+\frac{1}{2}\left[(\Omega^2-1)r^2-\frac{\nu^2}{r^2}\right]-Ae^{-l^2r^2}\bigg|_{R_1,R_2},
\end{equation}
and eliminating $\tilde{\mu}$ from Eq. (\ref{gg}) gives
\begin{equation}
\label{nu3}
\nu^2=R_1^2R_2^2\left[(1-\Omega^2)+\frac{2A}{X}e^{-l^2R_1^2}(e^{-l^2X}-1)\right],
\end{equation}
where $X$ is the `area' of the condensate and is again defined as $X\equiv R_2^2-R_1^2$. From integration of the normalisation condition (\ref{tor_norm}) between $R_1$ and $R_2$ it ensues that
\begin{eqnarray}
\frac{g}{\pi}&=&\frac{1}{4}(1-\Omega^2)X(R_1^2+R_2^2)-\frac{1}{2}(1-\Omega^2)R_1^2R_2^2\ln\left(\frac{R_2^2}{R_1^2}\right)\nonumber\\
&+&Ae^{-l^2R_1^2}(e^{-l^2X}-1)\left[R_1^2+\frac{1}{l^2}-\frac{R_1^2R_2^2}{X}\ln\left(\frac{R_2^2}{R_1^2}\right)\right]\nonumber\\
&+&AXe^{-l^2X}e^{-l^2R_1^2}.\nonumber\\
\label{nu1}
\end{eqnarray}

A further expression that connects the quantum of circulation and the inner and outer radii is readily obtained
from the minimisation of the free energy per particle, $F=E'-\mu
\bar{N}$, where $\bar{N}$ is the number of bosons in the condensate
\cite{fjs}. The following integral identity results:
\begin{equation}
\label{gg1}
g\Omega=2\nu\pi\int_{R_1}^{R_2}\frac{1}{r}\left[g|\psi|^2\right]dr.
\end{equation}
Integration of Eq. (\ref{gg1}) using Eq. (\ref{tf}), together with the expressions for $\tilde{\mu}$ and $\nu$ above, results in
\begin{equation}
    \begin{split}
\frac{g\Omega}{\pi\nu}&=\biggl[\frac{1}{2}(1-\Omega^2)(R_1^2+R_2^2)+Ae^{-l^2R_2^2}\\
&+\frac{AR_1^2}{X}e^{-l^2R_1^2}(e^{-l^2X}-1)\biggl]\ln\left(\frac{R_2^2}{R_1^2}\right)-X(1-\Omega^2)\\
&-Ae^{-l^2R_1^2}(e^{-l^2X}-1)+2A\int^{R_2}_{R_1}\frac{1}{r}e^{-l^2r^2}dr.
\label{int}
\end{split}
\end{equation}
In order to proceed it becomes important to estimate the last
integral  in Eq. (\ref{int}) according to the size of $l^2 r^2$. The
following sections detail two possible limits.

{\it{$l^2R_2^2$ small.}} If $l^2R_2^2$ is assumed to be small, then
it follows that $l^2R_1^2$ is small as well. Thus expanding the
exponential terms in Eq. (\ref{nu3}) up to terms of order $l^6R_2^6$
 gives
\begin{equation}
\nu^2\sim R_1^2R_2^2\left[(1-\Omega^2)(1-q)+Al^4R_2^2\right],
\label{temp1}
\end{equation}
where we have assumed that $R_2^2\gg R_1^2$,  which implies that we
can write $\exp(-l^2X)\sim\exp(-l^2R_2^2)$. In the case of an
annular condensate, we have that $q>1$, so that the first term in
Eq. (\ref{temp1}) is negative. Therefore in order for the right hand
side of Eq. (\ref{temp1}) to be positive, it is required that
\begin{equation}
l^2R_2^2>\frac{(q-1)(1-\Omega^2)}{Al^2}=2 \left(1-\frac 1 q\right).
\end{equation}

Expansion of Eq. (\ref{nu1}) to order $l^6R_2^6$ (again assuming that $R_2^2\gg R_1^2$) results in an expression for $R_2$,
\begin{equation}
\begin{split}
p&\sim\frac{1}{4}(1-q)l^4R_2^2\biggl[R_2^2\biggl(1+\frac{ql^2}{(1-q)}\biggl(\frac{R_2^2}{3}-R_1^2\ln(R_2^2)\biggr)\biggr)\\
&\qquad\qquad\qquad\qquad-2R_1^2\ln(R_2^2)\biggr]\\
&\sim\frac{l^4R_2^4(1-q)}{4}+\frac{ql^6R_2^6}{6},
\label{ihy}
\end{split}
\end{equation}
where we assume that $R_2^2\gg R_1^2\ln(R_2^2/R_1^2)\sim R_1^2\ln(R_2^2)$. Notice that (\ref{ihy}) is actually the same approximation for $R_2$ as calculated for the disk condensate in the TF approximation (see Eq. (\ref{r222})). Finally we can consider Eq. (\ref{int}). Expanding the exponential under the integral in powers of $l^2r^2$ and simplifying (\ref{int}) gives
\begin{equation}
\begin{split}
\frac{g\Omega}{\pi\nu}&\sim\frac{1}{2}(1-\Omega^2)R_2^2\left[\ln(R_2^2)-2\right]\\
&\qquad+A\biggl[2\ln(R_2^2)-l^2R_2^2\left(\ln(R_2^2)+1\right)\\
&\qquad+\frac{1}{2}l^4R_2^4\left(\ln(R_2^2)+\frac{1}{2}\right)\biggr]\\
&\sim\ln(R_2^2)\left[\frac{1}{2}(1-\Omega^2)R_2^2(1-q)+2A\right],
\label{ihy2}
\end{split}
\end{equation}
where we assume that $\ln(R_2^2/R_1^2)\sim\ln(R_2^2)\gg 1$. It
follows that
\begin{equation}
\label{irhy}
\nu\sim\frac{2g\Omega}{\pi\ln(R_2^2)}\left[R_2^2(1-\Omega^2)(1-q)+4A\right]^{-1},
\end{equation}
and thus the inner radius, $R_1$, is computed  from (\ref{temp1}) to be
\begin{equation}
\label{idy}
R_1\sim\frac{\nu}{R_2}\left[(1-\Omega^2)(1-q)+Al^4R_2^2\right]^{-1/2}.
\end{equation}
There are now three equations (\ref{ihy}), (\ref{irhy}) and
(\ref{idy}) for $R_2$, $\nu$ and $R_1$ that describe the annular
condensate in the TF regime with $l^2r^2$ small. These provide a comparison to the
 numerical values corresponding to the parameters of
Bretin et al. \cite{bssd}. A comparison for the case of
$\Omega=0.92$ and $\Omega=0.93$ between the numerics (c.f. Fig.
\ref{dali2}(g,h)) and analytical estimates is given in Table
\ref{dalit2}. The TF analytical expressions describing the radii of the inner and
outer boundary are reasonnably good. However the quantum of
circulation (Eq. (\ref{irhy})) is not providing a suitable estimate.
 Indeed $\nu$ is still small and the inconsistency could be as a result of the difficulty in numerically
 discriminating between vortices inside the condensate and
  inside the inner boundary.

\begin{table}
\caption{\label{dalit2}A comparison, for parameters $\{g,A,l\}=\{955.95,24.83,0.07\}$, for $\Omega$ such that the condensate is an annulus. The values of the inner radius $R_1$, outer radius $R_2$ and quantum of circulation $\nu$ are calculated by a TF analysis and are given by Eq.'s (\ref{idy}), (\ref{ihy}) and (\ref{irhy}) respectively. Numerical values are provided (subscripts $n$) as a comparison.}
\begin{ruledtabular}
\begin{tabular}{lcccccc}
$\Omega$&$R_1$&$R_{1n}$&$R_{2}$&$R_{2n}$&$\nu$&$\nu_n$\\
\hline
0.92 & 1.00 & 0.00 & 12.94 & 13.30 & 1.30 & 0\\
0.93 & 3.72 & 3.25 & 13.52 & 13.80 & 1.37 & 11\\
\end{tabular}
\end{ruledtabular}
\end{table}

We note that the Thomas Fermi approximation is justified because the maximum of
$g|\psi|^2$ is larger than 1 (numerically around 4).
  Larger values of $\Omega$ which are not
  reached by the experiments would be better described by a LLL
  regime. Nevertheless, in the TF approximation, we can still
  analyze an intermediate case.

  {\it $l^2R_1^2$ small} and {\it $l^2R_2^2$ large.}
If we take $l^2R_1^2$ small and $l^2R_2^2$ large then this implies
that $l^2X$ is also large. Taking Eq.'s (\ref{nu3}), (\ref{nu1})
and (\ref{int}) as the starting point we note immediately that
(\ref{nu3}) simplifies to
\begin{equation}
\nu^2=R_1^2R_2^2(1-\Omega^2), \label{pl}
\end{equation}
while Eq. (\ref{nu1}) simplifies to
\begin{eqnarray}
\frac{g}{\pi}&\sim&\frac{1}{4}(1-\Omega^2)R_2^4-\frac{1}{2}(1-\Omega^2)R_1^2R_2^2\ln(R_2^2)\nonumber\\
&\qquad&-Ae^{-l^2R_1^2}\left[R_1^2+\frac{1}{l^2}\right]\nonumber\\
&\sim&\frac{1}{4}(1-\Omega^2)R_2^2\left[R_2^2-2R_1^2\ln(R_2^2)\right]-AR_1^2\left(1+\frac{1}{l^2R_1^2}\right)\nonumber\\
&\sim&\frac{1}{4}(1-\Omega^2)R_2^4-\frac{A}{l^2}, \label{pl2}
\end{eqnarray}
where we have assumed that
\begin{eqnarray*}
R_2^2&\gg&R_1^2,\\
R_2^2&\gg&2R_1^2\ln(R_2^2),
\end{eqnarray*}
the first condition following directly from the assumption that $l^2R_1^2$ is small and $l^2R_2^2$ is large.
Furthermore Eq. (\ref{int}) simplifies as
\begin{eqnarray}
\frac{g\Omega}{\pi\nu}&\sim&\left[\frac{1}{2}(1-\Omega^2)R_2^2-\frac{2AR_1^2}{X}\right]\ln(R_2^2)-R_2^2(1-\Omega^2)+\nonumber\\
&\qquad&A+\frac{A}{e}-2A\ln(lR_1)-\frac{A}{l^2R_2^2}e^{-l^2R_2^2}\nonumber\\
&\sim&\frac{1}{2}(1-\Omega^2)R_2^2\left[\ln(R_2^2)-2\right]\nonumber\\
&\sim&\frac{1}{2}(1-\Omega^2)R_2^2\ln(R_2^2), \label{pl3}
\end{eqnarray}
since
\begin{equation}
\int^{R_2}_{R_1}\frac{1}{r}e^{-l^2r^2}dr\sim-\ln(lR_1)+\frac{1}{2e}-\frac{1}{2l^2R_2^2}e^{-l^2R_2^2},
\end{equation}
in the limits $l^2R_1^2$ small and $l^2R_2^2$ large. Thus we get an expression for the outer boundary,
from Eq. (\ref{pl2}), as
\begin{equation}
\label{y1}
R_2\sim\left[\frac{4}{(1-\Omega^2)}\left(\frac{g}{\pi}+\frac{A}{l^2}\right)\right]^{1/4},
\end{equation}
and from Eq. (\ref{pl3}) we get an expression for the quantum of circulation
\begin{equation}
\label{y2} \nu\sim\frac{2g\Omega}{\pi(1-\Omega^2)R_2^2\ln(R_2^2)},
\end{equation}
with $R_2$ found from Eq. (\ref{y1}). Notice how the value of $\nu$ gets large in the limit $\Omega\rightarrow1$.
Putting Eq.'s (\ref{y1}) and (\ref{y2}) into Eq. (\ref{pl}) we arrive at the expression for the inner boundary
\begin{equation}
\label{y3}
R_1\sim\frac{g\Omega R_2}{2\pi\ln(R_2^2)}\left[(1-\Omega^2)^{1/2}\left(\frac{g}{\pi}+\frac{A}{l^2}\right)\right]^{-1}
\end{equation}
again for $R_2$ given by Eq. (\ref{y1}).

Such a case is not reached experimentally and is just before the LLL
regime. One could perform similar computations assuming that
$l^2R_1^2$ is also large.

\section*{VII. CONCLUSION}
Motivated by the experiments of \cite{bssd}, we provide numerical and
analytical computations which describe the properties of a
condensate placed in a harmonic plus Gaussian trap. We have seen
that, however close $\Omega$ gets to the harmonic trapping
frequency, the condensate becomes a large annulus containing a
triangular vortex lattice, contrary to that seen for a condensate
with a quadratic plus quartic term where the width of the condensate
decreases. We estimate the circulation in the central hole which is
higher than that corresponding to a regular lattice in this region.
 Also in a Thomas Fermi approximation, when the rotational velocity
 $\Omega$ is not too close to the harmonic trapping frequency, we
 estimate the radii of the condensate and the circulation in the
 inside hole, in a way which is consistent with numerics.

\section*{ACKNOWLEDGMENTS}

We acknowledge support from the French ministry grant {\it
ANR-BLAN-0238, VoLQuan}.

\end{document}